\documentclass[aps,pra,twocolumn,10pt]{revtex4-1}

\usepackage[utf8]{inputenc}
\usepackage{amsmath}
\usepackage{stmaryrd}
\usepackage{color}
\usepackage{mathrsfs}
\usepackage{yfonts}
\usepackage{rotating}
\usepackage{graphicx}
\usepackage{amsfonts}
\usepackage{mathrsfs}
\usepackage{amsmath}
\usepackage[american]{babel}
\usepackage{wasysym}
\usepackage{slashed}
\usepackage{braket}
\usepackage{graphicx}
\usepackage[caption=false]{subfig}
\usepackage{dsfont}
\usepackage{amsthm}
\usepackage{youngtab}

\usepackage{natbib}
\PassOptionsToPackage{breaklinks}{hyperref}
\usepackage[colorlinks,allcolors=black]{hyperref}

\theoremstyle{definition}

\usepackage{cleveref}
\crefname{proposition}{Proposition}{Propositions}
\crefname{theorem}{Theorem}{Theorems}
\crefname{definition}{Definition}{Definitions}
\crefname{lemma}{Lemma}{Lemmas}
\crefname{figure}{Figure}{Figures}
\crefname{corollary}{Corollary}{Corollary}
\crefname{conjecture}{Conjecture}{Conjectures}
\crefname{section}{Section}{Sections}
\crefname{appendix}{Appendix}{Appendixes}
\crefname{observation}{Observation}{Observation}
\crefname{remark}{Remark}{Remark}
\crefname{example}{Example}{Examples}
\crefname{equation}{Eq.}{Eqs.}
\crefname{table}{Table}{Tables}

\usepackage{tikz}

\usepackage{amssymb}
   \makeatletter
     \renewcommand\@make@capt@title[2]{%
      \@ifx@empty\float@link{\@firstofone}{\expandafter\href\expandafter{\float@link}}%
       {\textbf{#1}}\@caption@fignum@sep#2\quad}%
     \makeatother
\makeatletter
\renewcommand{\fnum@figure}{\textbf{Figure~\thefigure}}

\usepackage{mathtools}

\newcommand{\g}[0]{\gamma}
\newcommand{\al}[0]{\alpha}
\newcommand{\be}[0]{\beta}

\newcommand{\e}[0]{\varepsilon}

\newcommand{\la}[0]{\lambda}

\newcommand{\si}[0]{\sigma}

\newcommand{\bs}[1]{\textbf{#1}}
\newcommand{\ea}[1]{\begin{align}#1\end{align}}
\newcommand{\eq}[1]{\begin{equation}#1\end{equation}}

\newcommand{\ma}[1]{\mathcal{#1}}

\usepackage{cancel}

\begin{document}


\title{Qubit Picture of Virtual Particles}
\author{Gon\c{c}alo M. Quinta}
\email{goncalo.quinta@tecnico.ulisboa.pt}
\affiliation{Instituto de Telecomunica\c{c}\~{o}es, Lisboa, Portugal}


\begin{abstract}

We show that virtual particles, despite being unobservable, can be described by quantum operators which can be interpreted under certain conditions as valid qubit quantum states. For a single virtual fermion, we prove that such a state is a separable mixed 2-qubit state with a well-defined finite temperature. For spin-1 virtual bosons, we find them to be associated to 4-qubit operators which can be interpreted as quantum states for some gauges. We also study the creation of virtual pairs of fermions, where the pair is shown to be associated to an entangled 4-qubit operator, and show the corresponding quantum circuit. Finally, we prove that renormalization does not structurally affect these results. These findings represent new connections between quantum field theory, quantum information and quantum thermodynamics.

\end{abstract}

\maketitle

\section{Introduction}

The concept of virtual particles is as old as perturbative calculations of Quantum Field Theory (QFT). When Feynman diagrams were first introduced as a tool to efficiently calculate scattering amplitudes, it became natural to use the term ``virtual particle'' as the name for an internal line in these diagrams. This was due to the fact that they represented momentum space propagators of the Dirac equation, hence meriting the term ``particle'', while at the same time violating the relativistic energy-momentum relation, thus being physically unobservable, or ``virtual''. The violation of relativity was balanced by the Heisenberg uncertainty principle, whereby the energy-momentum relation can be violated as long as it happens in a short enough amount of time. Nevertheless, since perturbative calculations in QFT can be performed without ever introducing Feynman diagrams, the question of whether virtual particles are actual particles of a mere mathematical tool is still a topic of debate \cite{Fox:2008, Valente:2011}.

Regardless of the physical meaning of virtual particles, their influence on the physical world is well-established. The most frequent example of this is the Lamb shift effect \cite{Lamb:1947}, whereby virtual electrons and positrons repeatedly interact with the electron in the Hydrogen atom for brief instants, resulting in an energy shift between two orbitals which would not exist otherwise. Another famous example is the Hawking effect \cite{Hawking:1975}, in which a pair of virtual particles is created near the horizon of a black hole, such that one of them is trapped inside the horizon and the other escapes outside. The latter effect is particularly supportive of the interpretation that virtual particles do in fact exist but are simply unobservable. In fact, quarks possess exactly this kind of nature and their existence is hardly questioned, mostly due to their ability to explain so well hadrons as a building block of matter. Nevertheless, an isolated quark has never been observed in Nature. The lack of knowledge regarding the physical reality of virtual particles is even more unfortunate given the critical role they play as carriers of all known fundamental forces. A better understanding of virtual particles would thus benefit every field of research where subatomic interactions are important, ranging from particle physics, nuclear physics and condensed matter physics.

Another tentative indication that virtual particles possess a richer uncovered structure is demonstrated by recent works \cite{Alba:2017,Araujo:2019}. These show that not only do particle interactions in QFT typically involve qubit degrees of freedom (usually in the form of helicities) but that they also entangle them. These interactions are mediated by virtual particles, leading inexorably to the conclusion that virtual particles, as part of the quantum dynamical evolution, are responsible for this entanglement by acting as quantum operators. This is a strong indication that virtual particles can be associated to specific qubit quantum operators and that the most fundamental ingredients of QFT must have an interpretation from the Quantum Information (QI) point of view.

In this work, we will develop on how virtual particles possess key quantum properties which have been overlooked in the literature since their inception. In particular, we show that virtual fermions can be described by mixed 2-qubit quantum states in momentum space with a finite temperature and study the entanglement present in the qubit degrees of freedom. We also show that virtual photons can be described as 4-qubit operators acting on two-particle states, which can be interpreted as density matrices depending on the gauge considered. We use these results to understand from a QI perspective the process of creation of a virtual pair of fermions from a physical photon, with the corresponding quantum circuit. We also prove that the quantum degrees of freedom of the created particles are always entangled. Finally, we show that the process of renormalization influences the quantum state of virtual fermion via a modification of its temperature, whereas the operator associated to the virtual photon remains physically the same. These results shed a light on the nature of virtual particles and provide a newfound bridge between particle physics, quantum information and quantum thermodynamics.

\section{The Qubit Operator of a Virtual Fermion}

We consider a virtual fermion in the Feynman diagrammatic sense, i.e. associated to the presence of the fermionic propagator in scattering amplitudes. The latter propagator is always of the form
\eq{\label{DF}
D_{F}(k) = \frac{1}{\slashed{k} - m} = \frac{\slashed{k} + m}{k^2-m}
}
where we use the slash notation $\slashed{k} = k^{\mu} \g_{\mu}$, $\mu$ are the spacetime indexes (assuming Einstein summation), $k^{\mu}$ is the real-valued 4-momenta vector of the particle, $m$ its mass and $\g_{\mu}$ are the Dirac matrices taken in the Dirac representation (using the standard sign conventions of \cite{Bjorken:1964}), i.e. $\g^0 \equiv \si_3 \otimes I_2$ and $\g^i \equiv i (\si_2 \otimes \si_i)$ where $\si_i$ are the Pauli matrices and $I_2$ is the identity matrix of dimension 2. The eigenvectors of $D_{F}(k)$ are $\check{u}^{1}(k) = W_k(1, 0, 0, 0)^T$, $\check{u}^{2}(k) = W_k(0, 1, 0, 0)^T$, $\check{v}^{1}(k) = W_k(0, 0, 1, 0)^T$ and $\check{v}^{2}(k) = W_k(0, 0, 0, 1)^T$, where we define the matrix $W_k$ as
\ea{\label{Wk}
W_k & = \frac{(k_0+M_k)I-k_i \g^i \g_0}{\sqrt{2M_k(k_0+M_k)}} \,,
}
where Latin indexes are used for spatial coordinates, $I$ is the identity matrix and $M_k$ is given by
\eq{
M_k = \sqrt{k^2} = \sqrt{k^2_0-k^2_1-k^2_2-k^2_3}\,.
}
The quantity $M_k$, denoted here by off-shell mass, is a generalization of the mass concept which in the case of an on-shell particle (i.e. $k^2=m^2$) reduces to the usual rest mass $m$. In fact, all basis eigenvectors reduce to the standard Dirac spinor solutions of the Dirac equation in the on-shell condition. The spinors $\check{u}^{s}(k)$ and $\check{v}^{s}(k)$ obey a generalized form of the Dirac equation in momentum space, given by $(\slashed{k} - M_k) \check{u}^{s}(k) = 0$ and $(\slashed{k} + M_k) \check{v}^{s}(k) = 0$, although as we shall see, one cannot build a physically consistent equivalent of the Dirac equation in position space. The eigenvalues of the propagator (\ref{DF}) associated to $\check{u}^{i}$ are given by $1/(M_k-m)$ while the ones related to $\check{v}^{i}$ are of the form $1/(M_k+m)$. Using the eigenbasis, one can show that the propagator in Eq.~(\ref{DF}) can be written as
\ea{\label{eigenDF}
D_{F}(k) & = \frac{1}{M_k-m} \sum_{s} \check{u}^{s}(k)\overline{\check{u}}^{s}(k) \nonumber \\ 
& \hspace{10mm} + \frac{1}{M_k+m} \sum_{s} \check{v}^{s}(k)\overline{\check{v}}^{s}(k)\,, 
}
where the Dirac adjoint $\overline{u} = u^{\dagger} \g_0 $ was used and the conjugate is taken only on the complex coefficients and not on the quantity $M_k$ (which can assume real or complex values). The form of Eq.~(\ref{eigenDF}) is extremely suggestive of a mixed state composed as the sum of four density matrices, each one associated to an eigenvector of Eq.~(\ref{DF}). However, this interpretation is not correct since $D_{F}(k)$ is not hermitian, as one can directly check. The lack of hermiticity comes from the right multiplication by $\g_0$, so if we multiply again by $\g_0$ from the right and adjust the multiplicative factor, we obtain the matrix
\ea{\label{rho}
\check{\rho}(k) & = \left(\frac{k^2-m^2}{4k_0}\right) D_{F}(k)\g_0 \\
& = \left(\frac{M_k+m}{4 k_0}\right) \sum_{s} \check{u}^{s}(k)\check{u}^{s \dagger}(k) \nonumber \\ 
& \hspace{10mm} + \left(\frac{M_k-m}{4 k_0}\right)  \sum_{s} \check{v}^{s}(k)\check{v}^{s \dagger}(k)
}
which is not only hermitian but has also trace equal to 1. We use again the inverted hat notation to remind that we are dealing with virtual particles. The quantity $\check{\rho}(k)$ is as relevant as the Feynman propagator, since they differ by a simple multiplication by $\g_0$. It is also worth noting that such a multiplication (from the right or left) naturally occurs in any QFT calculation, since the propagator is always multiplied by a Dirac adjoint spinor containing a $\g_0$. We can also see that the trace of $\check{\rho}^2(k)$ is given by
\eq{
\textrm{Tr}[\check{\rho}^2(k)] = \frac{1}{4}\left(1+r^2_k\right)\,, \label{trace}
}
which lies in the range $[1/4,\infty[$, where we have defined the quantity 
\eq{\label{rk}
r_k = \frac{\sqrt{k^2_1+k^2_2+k^2_3+m^2}}{k_0}\,.
}
Finally, the eigenvalues of $\check{\rho}(k)$ can be written as
\eq{\label{eigenvalues}
\lambda_{\pm} = \frac{1}{4}\left(1\pm r_k\right)\,,
}
each one having a multiplicity of 2. It is only natural to inquire under which conditions does $\check{\rho}(k)$ stand as a valid density matrix. For this to happen, the matrix needs to be positive semi-definite which amounts to requiring that all eigenvalues are equal or larger than zero. From Eq.~(\ref{eigenvalues}), one concludes that this happens for $|k_0|\geq \sqrt{k^2_1+k^2_2+k^2_3+m^2}$ or equivalently,
\eq{\label{physicalrk}
|r_k| \leq 1 \,,
}
which also implies that $\textrm{Tr}[\check{\rho}^2(k)] \leq 1/2$, by virtue of Eq.~(\ref{trace}). The condition (\ref{physicalrk}) can also be stated as
\eq{
|M_k| \geq m\,,
}
i.e. whenever the off-shell mass is larger in magnitude than the on-shell mass $m$. In other words, a virtual fermion can only have a physically valid density matrix whenever it has enough energy to at least create a real fermion of the same species. Under condition (\ref{physicalrk}) then, the matrix $\check{\rho}(k)$ represents a valid $4 \times 4$ mixed density matrix. Since $\check{\rho}(k)$ is obtained directly from the propagator, we consider it to be the associated density matrix of a virtual fermion, when in the regime (\ref{physicalrk}). On the other hand, the fact that $\check{\rho}(k)$ is only physically consistent for certain values of momentum $k$ implies that a representation in position space, as the associated Fourier transform, cannot be constructed due to the integration over all momenta. As a consequence, a generalization of the Dirac equation in position space for a virtual fermion is not feasible without relaxing the condition of semi-positiveness for the density matrix of a quantum state. This is expected at some level, however, since it would be peculiar to have universally valid quantum states for physically unobservable entities. In fact, the $|M_k| \geq m$ sector is only relevant from an interpretative point of view. Whether the momentum in $\check{\rho}(k)$ is being integrated (e.g. in a loop) or not (e.g. in a simple scattering), how we interpret the quantity $\rho(k)$ is of no physical consequence. What matters is its quantum information properties.

The fact that $\check{\rho}(k)$ could be interpreted as a physically valid $4 \times 4$ density matrix suggests that it might also be associated to a 2-qubit system. This motivates the introduction of a notation more familiar to QI, by introducing the qubit notation $\{\ket{00} = (1,0,0,0)^T,\ket{01} =(0,1,0,0)^T,\ket{10}=(0,0,1,0)^T,\ket{11}=(0,0,0,1)^T\}$. To find the physical interpretation of these qubits, we must first define the energy and spin projection operators in the rest-frame, respectively denoted by $\Lambda(\lambda_E)$ and $\Sigma(\lambda_s)$, with explicit form given by
\ea{
\Lambda(\lambda_E) & = \frac{1+(-1)^{\lambda_E} \g_0}{2} \\
\Sigma(\lambda_s) & = \frac{1 -(-1)^{\lambda_s} \g_5 \g_3 \g_0}{2}
}
with $\lambda_E \in \{0,1\}$, $\lambda_s \in \{0,1\}$ and $\g_5 = i \g_0\g_1\g_2\g_3$ \cite{Bjorken:1964}. These operators are essential for attributing physical meaning to each Dirac spinor. The values $\lambda_E=0,1$ (which we shall denote as particle type values) correspond to particle and anti-particle solutions, respectively. One could also note that the energy-projection operators have exactly the same form as the charge projection operators, so the first qubit can also be interpreted as charge sign. 
The values $\lambda_s=0,1$ are correspondingly associated to spin up and spin down solutions. Using the qubit braket notation, this readily leads to the relation
\eq{
\Lambda(\lambda_E)\Sigma(\lambda_s) = \ket{\lambda_E \, \lambda_s}\bra{\lambda_E \, \lambda_s}\,
}
from which we conclude that the qubits associated to $\check{u}^{s}(k)$ and $\check{v}^{s}(k)$ can be physically identified with particle/anti-particle and spin up/down, i.e. we can write the eigenbasis of Eq.~(\ref{DF}) more succinctly in the Dirac notation as
\eq{
\ket{\lambda_E,\lambda_s;k} \equiv W_k \ket{\lambda_E \, \lambda_s}\,.
}
One can now insert this in Eq.~(\ref{rho}), resulting in
\eq{\label{rhoqs}
\check{\rho}(k) = \sum_{\lambda_E,\lambda_s} \frac{1}{4}\left(1+(-1)^{\lambda_E}\frac{m}{M_k}\right) \check{\rho}(\lambda_E,\lambda_s;k)\,,
}
where
\eq{\label{rho1fermion}
\check{\rho}(\lambda_E,\lambda_s;k) = \frac{M_k}{k_0} \ket{\lambda_E, \lambda_s; k} \bra{\lambda_E, \lambda_s; k}
}
corresponds to the individual pure density matrix of the virtual fermion with particle type $\lambda_E$ and spin $s$. It is straightforward to check that $\textrm{Tr}[\check{\rho}(\lambda_E,\lambda_s;k)] = 1$ and that $\check{\rho}(\lambda_E,\lambda_s;k)$ is positive-definite precisely in the region defined by $|r_k|\leq 1$. The factors $p_{E,s} \equiv \left(1+(-1)^{\lambda_E}\frac{m}{M_k}\right)/4$ obey $0 \leq p_{E,s} \leq 1$ for $|r_k|\leq 1$, so they can safely be interpreted as the probabilities associated to the mixture of density matrices $\check{\rho}(\lambda_E,\lambda_s;k)$.

The form (\ref{rhoqs}) highlights the separation between particle type, spin and kinematic degrees of freedom contained in each spinor solution of Eq.~(\ref{DF}). As a consequence, one can attribute a physical meaning to the operations of partial trace and partial transpose for the density matrix $\check{\rho}(k)$, which in turn allows the study of entanglement between these quantum degrees of freedom.

It's important to note that, while the case $r_k>1$ does not allow a quantum state interpretation for $\check{\rho}(k)$, it does not change the fact that the virtual fermion propagator is still a quantum operator acting on qubits with a physical meaning. The condition $r_k>1$ simply results in a density matrix with negative probabilities, which is of no physical consequence since the operator is always acted on both sides by external states. This situation is similar to the use of Wigner functions in quantum mechanics, which correspond to probability distributions which can assume negative values that actively play a role in quantum dynamics in phase space. The negative probabilities are harmless since they are integrated out whenever an observable is calculated. As an object directly obtained from the propagator (\ref{DF}), $\check{\rho}(k)$ is related to the probability amplitude for a particle to go from one point to another in momentum space. The fact that it contains negative eigenvalues is inconsequential since observables obtained from it have integrations over all momenta.

\section{Entanglement and Temperature of a Virtual Fermion}

As it has already become apparent from Eqs.~(\ref{trace}) and Eqs.~(\ref{eigenvalues}), the quantum information properties of a virtual fermion are essentially dependent on the parameter $r_k$ defined in Eq.~(\ref{rk}). This is especially clear when considering that the eigenvalues of $\check{\rho}(k)$, which are the main building blocks for many quantities in quantum information theory, are also a function of $r_k$ alone.

To see this, let's begin by analyzing the entanglement properties of $\check{\rho}(k)$. Since the latter state is a $2\times2$ qubit system, a standard Positive Partial Transpose (PPT) test \cite{Horodecki:1996} is necessary and sufficient to determine if the state is entangled or not. We thus partially transpose any of the qubits and look at the signs of the resulting eigenvalues. If any of the eigenvalues is negative, the state is entangled, otherwise it's separable. Partially transposing any of the qubits leads to a matrix whose eigenvalues are the exact same values $\lambda_{\pm}$ of Eq.~(\ref{eigenvalues}). Since $|r_k|\leq 1$, we have that $\lambda_{\pm} \geq 0$ and so $\check{\rho}(k)$ is always separable in the regime where it is well-defined as a density matrix.

Although it is frequently the case that the state of a single particle is separable, there are cases where it can be entangled, such as the case of a flavor oscillating neutrino \cite{Blasone:2009, Kayser:2010}, for example. In fact, the individual states $\check{\rho}(\lambda_E,\lambda_s;k)$ are almost always entangled themselves. To see this, it's easier to quantify the entanglement in a single scalar quantity, which in this case we will choose as the concurrence $C(\rho)$ \cite{Coffman:2000}. For a general pure state $\ket{\psi} = a\ket{00}+b\ket{01}+c\ket{10}+d\ket{11}$, the concurrence has the closed form $C(\psi) = 2|ad-bc|$. Now, the density matrices $\check{\rho}(\lambda_E,\lambda_s;k)$ are associated to the normalized pure states $\sqrt{\frac{M_k}{k_0}} \ket{\lambda_E, \lambda_s; k}$ which will always have one of the components equal to 0. For example, the case $(\lambda_E,\lambda_s)=(0,0)$ gives the state $\sqrt{\frac{k_0+M_k}{2k_0}}\ket{00}+\frac{k_3}{\sqrt{2k_0(k_0+M_k)}}\ket{10}+\frac{k_1+i k_2}{\sqrt{2k_0(k_0+M_k)}}\ket{11}$ with an associated concurrence equal to $|\sqrt{(k^2_1+k^2_2)/k_0}|$. Calculating all instances of particle type and spin, one concludes that the general expression for the concurrence is of the form $C(\check{\rho}(\lambda_E,\lambda_s;k)) = \left|\sqrt{(k^2_1+k^2_2)/k_0}\right|$. The fact that the concurrence does not depend on $k_3$ is a peculiarity of the Dirac representation used for the $\g^{\mu}$ matrices. It is a direct consequence of the implicit assumption in the Dirac representation that the spin is oriented in the same axis as the vector $(0,0,1)$. This is also apparent from the form of the spin projector operator $\Sigma(\lambda_s)$, which clearly singles out the direction aligned with the $z$ component. The frame-independent interpretation of $k^2_1+k^2_2$ is the norm of the transverse momentum vector $\bs{k}_T = (k_1,k_2,0)$, where a bold symbol is used for the spatial components of 4-vectors. A representation independent formula for the concurrence is thus
\eq{
C(\check{\rho}(\lambda_E,\lambda_s;k)) = \frac{||\bs{k}_T||}{\left|\sqrt{k_0}\right|}\,,
}
implying that the entanglement between the particle type and spin degrees of freedom of $\check{\rho}(\lambda_E,\lambda_s;k)$ vanishes whenever there is no transverse component of the momentum.

Knowing that the state of a virtual fermion is mixed leads to the natural question of whether it is a thermal ensemble or not. This is actually the case, as one can show that the density matrix $\check{\rho}(k)$ can be written as a single exponential in the form (proof in Section~A of the Appendix)
\eq{\label{rhoMt}
\check{\rho}(k) = \frac{e^{-\be H}}{\textrm{Tr}[e^{-\be H}]}
}
where $H = -\frac{1}{r_k}(m\g^0-k_i \g^i \g^0)$ is the Hamiltonian in the momentum space and 
\eq{\label{betaV}
\be = \frac{1}{2k_0}\ln\left(\frac{1+r_k}{1-r_k}\right)
}
is the inverse temperature associated to the thermal state, valid for all $r_k$. For $|r_k| \leq 1$, $\be$ is real and positive, while for $|r_k|>1$ the temperature becomes complex. In the later case, virtual fermions have an inverse temperature of the form $\be = \be_r + i \be_i$ and so $\check{\rho}(k)$ acquires a unitary component $e^{-i \be_i H}$ representing an evolution in momentum space during a time $\be_i$ under the Hamiltonian $H$. This sharp difference in behavior between lower energy ($|k_0|<m$) and higher energy ($|k_0|>m$) virtual fermions is a reflection of the positivity properties of the eigenvalues in Eq.~(\ref{eigenvalues}). Finally, note that $\be$ diverges in the on-shell limit $r_k \to 1$, i.e. the temperature goes to 0 for a real particle, as expected.

\section{The Qubit Operator of a Virtual Photon}

Having analysed the quantum information properties of a virtual fermion, the remaining task is to consider higher spin particles. We shall focus on the spin 1 case, since it exhausts all possible higher spins for QED.

The spin 1, or photon, case is less straightforward to derive. The photon propagator $(-g_{\mu\nu} + (\xi-1) \frac{k_{\mu}k_{\nu}}{k^4} )/k^2$ involves only spacetime indices, so by itself it cannot have a qubit interpretation similar to the fermionic one. One may, however, derive an analogous object to the Dirac propagator by writing, for example, the electron pair to muon pair scattering $e^+ e^- \to \mu^+ \mu^-$ in the form (derivation in Section B of the Appendix)
\ea{\label{Memu}
\ma{M}_{e^+ e^- \to \mu^+ \mu^-} & = \nonumber \\
& \hspace{-20mm} \frac{e^2}{p^2} \textrm{Tr}\left[(I \otimes u^{s_3}(p_3) v^{s_4 \dagger}(p_4)) D_{\g}(k)  (u^{s_2}(p_2) v^{s_1 \dagger}(p_1) \otimes I)\right]
}
where $k=p_1+p_2$, $D_{\g}(k)$ is an operator of the form
\eq{\label{photonOp}
D_{\g}(k) = \sum_{i} \Xi^{i}(k) \otimes \Xi^{i \dagger}(k)
}
and
\ea{\label{Xi}
\Xi^{i}(k) & = \slashed{\e}^{(i)}(k) \g^0 \nonumber \\
& = M_{\e^{(i)}(k)} \sum_{i_1,i_2}\ket{i_1,i_2;\e^{(i)}(k)}\bra{i_1,i_2;\e^{(i)}(k)}
}
where $i$ labels the basis vectors of photon polarization vectors $\e^{(i)}(k)$. The vectors $\ket{n,l;\e^{(i)}}$ form an eigenbasis of the operator $\slashed{\e}^{(i)}(p) \g_0$ and, similarly to spinors, are of the form ${\ket{l,n;\e^i(k)} = W_{\e^{i}(k)} \ket{l, n}}$ ($l,n = 0,1$), where the momentum 4-vector in Eq.~(\ref{Wk}) is substituted by the polarization vector $\e^{i}_{\mu}(k)$. The clear operator separation between ingoing and outgoing spinors, motivates the interpretation of $D_{\g}(k)$ as the operational object associated to a virtual photon and the decomposition (\ref{Xi}) shows that it makes sense to attribute a qubit interpretation to the action of a virtual photon.

Although the vectors $\ket{l,n;\e^i(k)}$ share the same form as the eigenvectors in the spin 1/2 case, the virtual photon operator is fundamentally different due to the presence of gauge freedom in the polarization vectors. For example, we have that $\textrm{Tr}[D_{\g}(k)]=|\sum_i \e^{(i)}_0|^2$, which we could use to normalize $D_{\g}(k)$ to have trace equal to 1. However, in gauges where $\e_0(k) = 0$ (which are the most widely used) we would have singular matrix elements for $D_{\g}(k)$. In addition, the eigenvalues of $D_{\g}(k)$ would change sign depending on the gauge, thus the positive semi-definiteness of $D_{\g}(k)$ is not gauge invariant. This same reasoning applies to the associated temperature (\ref{betaV}) one would obtain by following the same steps used to derive (\ref{rhoMt}), i.e. the temperature would not have physical meaning since it would not be gauge invariant. As a consequence, $D_{\g}(k)$ can only be interpreted as a quantum state of a virtual photon under gauges which satisfy the valid density matrix requirements. A particular gauge example which satisfies positive semi-definiteness is for real representations of $\e^{(i)}_{\mu}$ such that $\e^{(i)}_{0}>0$ and $\e^{(i)}_{0}>\sqrt{\left(\e^{(i)}_{1}\right)^2+\left(\e^{(i)}_{2}\right)^2+\left(\e^{(i)}_{3}\right)^2}$ (proof in Section B of the Appendix).

Despite the shortcomings in interpreting $D_{\g}(k)$ as a quantum state, it provides an interpretation of how a virtual photon interacts with the external spinors from the qubit point of view. The $D_{\g}(k)$ represents an operator that acts on a state of two spin 1/2 particles, or equivalently, a 4-qubit operator. In the case of the $e^+ e^- \to \mu^+ \mu^-$ interaction in Eq.~(\ref{Memu}), the virtual photon operator acts on the first two particles by combining them into the state $(u^{s}(p_2) v^{s \dagger}(p_1) \otimes I)$, followed by an action on the state of the final two particles, in the form of $(I \otimes u^{s}(p_3) v^{s \dagger}(p_4))$.

The quantity $\Xi(k)$ in Eq.~(\ref{Xi}) also appears naturally by itself in different contexts. For example, is it straightforward to show (c.f. Section B of the Appendix) that the scattering amplitude of an electron with a photon can be written in the form
\eq{
\ma{M}_{e^- \g \to e^- \g} = \frac{4e^2 k_0}{s-m^2} u(p_3)^{\dagger} \Xi^{\dagger}(p_4) \check{\rho}(k) \Xi^{\dagger}(p_2) u(p_1)
}
where $k=p_1+p_2$ and $s$ is the center of mass energy. In this case, $\Xi(k)$ appears multiplied by a spinor $u(p)$, such that the resulting object $\Xi(k)u(p)$ is also a spinor. This is consistent with the subsequent multiplication by $\check{\rho}(k)$, representing an object with spin 1/2, the virtual fermion. This contrasts with the case of $e^+ e^- \to \mu^+ \mu^-$, where the initial state is a tensor product of two spinors (two 2-qubit states) instead and thus the virtual photon operator $D_{\g}(k)$ must be a 16 dimensional matrix (4-qubit operator) for consistency. It's also worth mentioning that $k^2>p^2_1=m^2$, which implies that $\check{\rho}(k)$ is positive semi-definite in this case. In fact, any scattering involving only one virtual fermion and external physical particles will always result in a positive semi-definite $\check{\rho}(k)$, since the sum of the external momenta will always have a norm greater than the square of the rest mass of any one of the particles.

\section{The Qubit Picture of Virtual Pair Creation}

Having access to a qubit interpretation for virtual fermions and photons, it's interesting to understand how the creation of a pair of virtual fermions can fit into the qubit picture. By making use of Eq.~(\ref{rhoqs}), one can write the 1-loop scattering amplitude in the form (proof in Section C of the Appendix)
\eq{\label{1loop}
\ma{M}_{1-\textrm{loop}} = -\frac{e^2}{16} \textrm{Tr}\left[\check{\rho}_{2F}(p)(\Xi(p) \otimes \Xi^{\dagger}(p))\right]
}
where 
\eq{\label{rho2F}
\check{\rho}_{2F}(p) = R(p) S
}
and
\ea{
R(p) & = \int \frac{d^4k}{(2\pi)^4}  \, \check{\rho}(p-k) \otimes \check{\rho}(k) \label{Rgate} \\
S & = \sum_{\substack{\la_{E},\la_{s}\\ \la_{E'},\la_{s'}}} \ket{\la_{E'},\la_{s'}}\ket{\la_{E},\la_{s}} \bra{\la_{E},\la_{s}}\bra{\la_{E'},\la_{s'}}\,. \label{Sgate}
}
A number of comments are in order. First of all, the form (\ref{1loop}) explicitly separates between objects related to ingoing and outgoing states and dynamical evolution in-between, represented by $\check{\rho}_{2F}(k)$. The physical setup is depicted in Figure 1. The latter quantity is a 4-qubit operator constructed as a mixture of the two 2-qubit states related to the two virtual fermions created in the loop, as evident by the presence of the matrices $\check{\rho}(p-k)$ and $\check{\rho}(k)$. A direct inspection of Eq.~(\ref{Sgate}) shows that the matrix $S$ can be interpreted as a SWAP quantum gate interchanging the particle type qubits $\la_{E}$ and $\la_{E'}$, as well as the spin qubits $\la_{s}$ and $\la_{s'}$, such that $(A \otimes B)S = S (B \otimes A)$. This interchanging of qubits entangles the separable matrix $R(p)$, such that the operator $\check{\rho}_{2F}(k)$ is always entangled. It is also tentative to interpret $\check{\rho}_{2F}(k)$ as an entangled 4-qubit density matrix, although it is possible to show that such an interpretation is never possible (proof in Section C of the Appendix).
\begin{figure}[h!]
\centering
\includegraphics[width=0.48\textwidth]{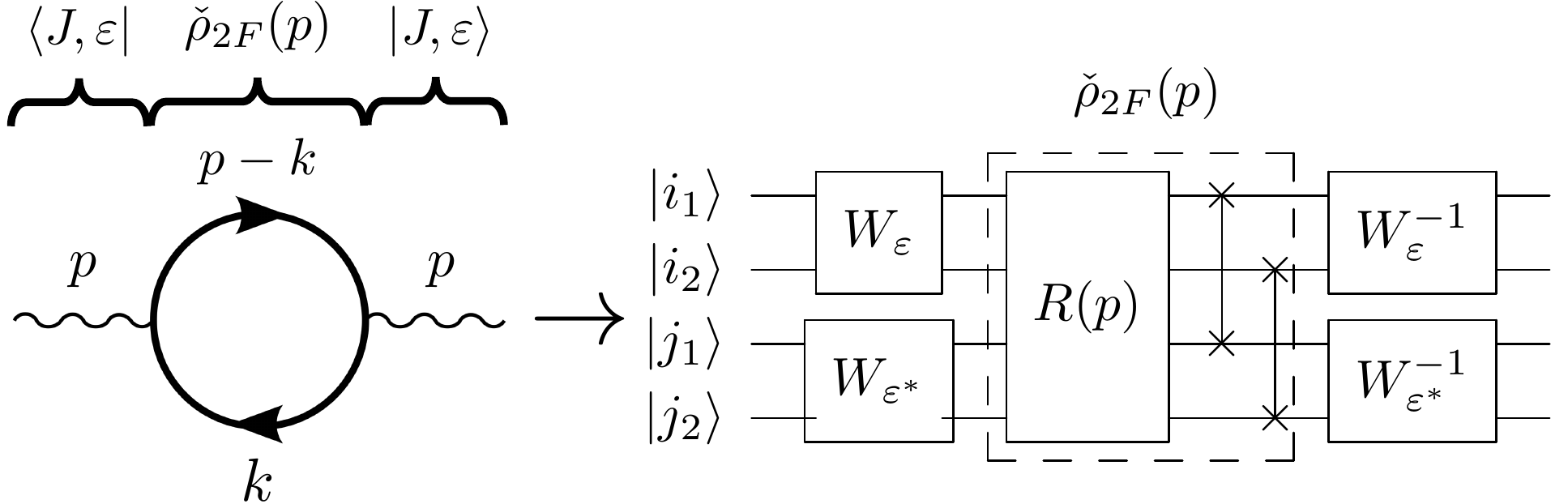}
\caption{On the left: the 1-loop scattering amplitude written in terms of qubit states and operators. The incoming and outgoing real photons are associated to the 4-qubit states $\ket{J,\e} \equiv \ket{i_1,i_2; \e(p)}\otimes\ket{j_1,j_2; \e^{*}(p)} = W_{\e} \ket{i_1,i_2} \otimes W_{\e^{* }} \ket{j_1,j_2}$, while the pair of virtual fermions is represented by $\check{\rho}_{2F}(k)$. On the right, the equivalent quantum circuit construction of the amplitude $\bra{J,\e}\check{\rho}_{2F}(k)\ket{J,\e}$ prior to measurement of the outgoing qubits. Taking an average over all measurements for all ingoing and outgoing qubits results in the total amplitude for the 1-loop scattering.}
\label{virtual_pair}
\end{figure}
Finally, by using standard integrals in QFT, the form (\ref{rho2F}) allows a straightforward decomposition $\check{\rho}_{2F}(k) = \check{\rho}^{0}_{2F}(k) + \check{\rho}^{\textrm{div.}}_{2F}(k)$ into a finite term $\check{\rho}^{0}_{2F}(k)$ plus a divergent piece $\check{\rho}^{\textrm{div.}}_{2F}(k)$. We refrain from showing the explicit forms since they do not add any new intuition for this work. The qubit point of view in Eq.~(\ref{1loop}) can also be understood more intuitively by using Eq.~(\ref{Xi}), resulting in
\eq{\label{1loopnew}
\ma{M}_{1-\textrm{loop}} = -e^2 |\e_0(p)|^2 \braket{\check{\rho}_{2F}(p)}_{\e}
}
where we define
\eq{
\braket{\check{\rho}_{2F}(p)}_{\e} = \frac{\sum_{J} \bra{J,\e} \check{\rho}_{2F}(p) \ket{J,\e}}{\sum_{J} \braket{J,\e|J,\e}}
}
and use the shorthand notation {$\ket{J,\e} \equiv \ket{i_1,i_2; \e(p)}\otimes\ket{j_1,j_2; \e^{*}(p)}$}, with $J=(i_1,i_2,j_1,j_2)$. The quantity $\braket{\check{\rho}_{2F}(p)}_{\e}$ corresponds to an expectation value over all qubits of the initial and final photons. In particular, each term $\bra{J,\e} \check{\rho}_{2F}(p) \ket{J,\e}$ can be associated to the probability amplitude of an initial photon, described by the state $\ket{J,\e}$, being projected by the 2-fermion operator $\check{\rho}_{2F}(p)$ and finally collapsed again to the initial state. Since, by Eq.~(\ref{Xi}), the initial photon is a uniform mixture of all qubits $J$, the amplitude $\ma{M}_{1-\textrm{loop}}$ is thus the average over all such possibilities for each set of qubits. Note that the multiplication by $|\e_0(p)|^2$ in Eq.~(\ref{1loopnew}) poses no problem as it is cancelled by the term $\sum_{J} \braket{J,\e|J,\e} = 16|\e_0(p)|^2/(M_{\e(p)}M_{\e^*(p)})$. The result in Eq.~(\ref{1loopnew}) makes it straightforward to find the quantum circuit which simulates the pair creation process, represented in Figure 1.

\section{Renormalization}

It is natural at this point to inquire what are the consequences of the inevitable renormalization that must take place in the virtual particle propagators. We begin with the fermion case, for which a renormalized propagator will always have the form $D^{R}_F(p) = 1/(\slashed{p}-m+\Sigma_R(\slashed{p}))$ where $\Sigma_R(\slashed{p})$ is a finite matrix correction. Writing $\Sigma_R(\slashed{p}) = A_1 m + A_2 \slashed{p}$, where $A_i$ are finite coefficients, and inserting this in the renormalized propagator, leads to the conclusion that the operator $\check{\rho}(k)$ from Eq.~(\ref{rho}) simply suffers a transformation of the form $m \to (m+A_1)/(1+A_2)$ when renormalization is taken into consideration. This means that all results obtained for a virtual fermion are still valid, subject only to a finite mass scaling.

The case of virtual photon is much simpler and it can be straightforwardly shown that the overall operator $D_{\g}(k)$ is multiplied by a constant while the gauge is divided by it (proof in Section D of the Appendix). Consequently, nothing changes from the point of view of the qubit structure of $D_{\g}(k)$.


\section{Conclusions}

We demonstrated how virtual particles, despite being physically unobservable, can be associated to quantum operators acting on qubit states. In addition, whenever the off-shell mass is larger than the rest mass in magnitude, the fermion propagator can be interpreted as a well-defined density matrix in momentum space. We showed that this state can be interpreted a mixed 2-qubit thermal density matrix, with one qubit associated to particle type (or charge) and another one to spin. It was proven that although these two quantum degrees of freedom exist in a separable global state, the individual density matrices composing the system's state are actually entangled in most cases. Virtual photons were also found to be associated to quantum operators, in this case acting on 4-qubit states. However, the lack of gauge invariance limits the interpretation of these operators as quantum states, possible only under certain gauges. The operators for virtual fermions and photons were used to understand how the process of virtual fermionic pair creation can be understood from the qubit point of view, where it was proven that the virtual pair is associated to an entangled operator. Finally, it was proven that renormalization has a simple effect on the virtual fermion operators in the form of a mass scaling, while having no physical effect on the virtual photon operators.

Overall, the observation that virtual particles have well-defined quantum properties sometimes similar to real particles is suggestive that virtual particles may have a physical existence despite their inability to be observed in Nature, akin to quarks. Furthermore, the fact that the operators associated to virtual fermions and bosons have a natural qubit interpretation is also indicative not only are Quantum Field Theory and Quantum Information related at the fundamental but also that new unexplored connections may exist between those fields.\\


\textit{Acknowledgements} -- The author thanks the support from Funda\c{c}\~{a}o para a Ci\^{e}ncia e a Tecnologia (Portugal), namely through project CEECIND/02474/2018 and project EXPL/FIS-PAR/1604/2021 QEntHEP - Quantum Entanglement in High Energy Physics.


\clearpage
\pagebreak
\newpage

\begin{center}
\textbf{\large Apendix for ``The Qubit Picture of Virtual Particles''}
\end{center}
\setcounter{equation}{0}
\setcounter{figure}{0}
\setcounter{table}{0}
\setcounter{page}{1}
\makeatletter
\renewcommand{\theequation}{A\arabic{equation}}
\renewcommand{\thefigure}{A\arabic{figure}}
\renewcommand{\bibnumfmt}[1]{[A#1]}
\renewcommand{\citenumfont}[1]{A#1}

\section*{Section A: Proof of Eq.~(19)}

In this section we shall demonstrate a proof of Eq.~(19). Starting from Eq.~(4), we expand $\check{\rho}(k)$ as
\eq{
\check{\rho}(k) = \frac{1}{4}I + \frac{1}{4k_0}(m \g^0 - k_i \g^i\g^0)\,.
}
Now we define the matrix $B$ by
\eq{
B = \frac{m \g^0 - k_i \g^i\g^0}{\sqrt{m^2+k^2_1+k^2_2+k^2_3}} = \frac{m \g^0 - k_i \g^i\g^0}{k_0 r_k}
}
which obeys the relation $B^2=I$. We then introduce the hyperbolic angle $\al$ via the relations
\ea{
\cosh(\al) & = \frac{\textrm{sign}(1-r_k)}{\sqrt{1-r^2_k}}\,, \label{cosh} \\
\sinh(\al) & = \frac{\textrm{sign}(1-r_k) r_k}{\sqrt{1-r^2_k}}\,, \label{sinh}
}
which obey the required condition $\cosh^2(\al)-\sinh^2(\al)=1$. Writing $\check{\rho}(k)$ using the hyperbolic angle and the matrix $B$, we obtain
\ea{
\check{\rho}(k) & = \textrm{sign}(1-r_k) \frac{\sqrt{1-r^2_k}}{4}\left(\cosh(\al)I + \sinh(\al) B\right)  \\
& = \textrm{sign}(1-r_k) \frac{\sqrt{1-r^2_k}}{4} e^{M}\,. \label{rhomid}
}
where $M = \al B$. We can now use Eq.~(\ref{cosh}), together with the relation \cite{Gradshteyn:2014}
\eq{\label{coshm1}
\cosh^{-1}(z) = \ln(z+\sqrt{z-1}\sqrt{z+1})
}
valid for all complex $z \in \mathbb{C}$ with a branch cut in $]-\infty,1]$, to conclude that
\eq{
\al = \frac{1}{2}\ln\left(\frac{1+r_k}{1-r_k}\right)\,,
}
which implies that
\eq{
M = \frac{1}{2k_0 r_k }\ln\left(\frac{1+r_k}{1-r_k}\right) (m \g^0 - k_i \g^i\g^0)\,.
}
Finally, we have that
\eq{
\textrm{Tr}[e^M] = \textrm{Tr}[\cosh(\al)I + \sinh(\al) B] = \frac{4 \, \textrm{sign}(1-r_k)}{\sqrt{1-r^2_k}}
}
which, along with Eq.~(\ref{rhomid}), implies that
\eq{\label{rhoM2}
\check{\rho}(k) = \frac{e^M}{\textrm{Tr}[e^M]}\,.
}
On the other hand, a thermal ensemble of 1-particle quantum states is always of the form $\rho = e^{-\be H}/(\textrm{Tr}[e^{-\be H}])$ where $H$ is the system's Hamiltonian and $\beta=1/T$, with $T$ being the associated temperature. Comparing this with Eq.~(\ref{rhoM2}), we see that they are of the same form. The remaining task is to determine the temperature. This can be readily found by noting that whatever the Hamiltonian is, its eigenvalues will always be the virtual fermion energies $k_0$ and $-k_0$, each with multiplicity 2. Using the latter fact and decomposing the Hamiltonian in the diagonal basis, results in $\textrm{Tr}[e^{-\be H}] = 4 \cosh(k_0 \be)$. On the other hand, since $\textrm{Tr}[e^{M}] = \textrm{Tr}[e^{-\be H}] = 4 \textrm{sign}(1-r_k) (1-r^2_k)^{-1/2}$, we can use again Eq.~(\ref{coshm1}) to arrive at
\eq{\label{beta}
\be = \frac{1}{2k_0}\ln\left(\frac{1+r_k}{1-r_k}\right)\,.
}
Since $M=-\be H$, this implies that the associated Hamiltonian in the momentum space must be
\eq{
H = -\frac{1}{r_k}(m\g^0-k_i \g^i \g^0)\,,
}
which, as expected, has the eigenvalues $k_0$ and $-k_0$, each with multiplicity 2. This concludes the proof.

\section*{Section B: Proofs of virtual photon identities}

Following the usual Feynman rules for QED, the scattering amplitude for the s-channel interaction $e^+ e^- \to \mu^+ \mu^-$ is given by \cite{Schwartz:2013}
\ea{
\ma{M}_{e^+ e^- \to \mu^+ \mu^-} & = -\frac{ie^2}{p^2} [v^{\dagger s_2}(p_2)\g^{0}\g^{\mu}u^{s_1}(p_1)] \nonumber \\
& \left(-g_{\mu\nu}+(1-\xi)\frac{p_{\mu}p_{\nu}}{p^2}\right) [u^{\dagger s_3}\g^{0}(p_3)\g^{\mu}v^{s_4}(p_4)]
}
A straightforward manipulation using the identities $\textrm{Tr}[v^{\dagger}Au] = \textrm{Tr}[(uv^{\dagger} \otimes I)(A \otimes I)]$, $\textrm{Tr}[A \otimes B] = \textrm{Tr}[A]\textrm{Tr}[B]$ and $\textrm{Tr}[AB]=\textrm{Tr}[BA]$ leads to the form
\ea{
\ma{M}_{e^+ e^- \to \mu^+ \mu^-} & = -\frac{ie^2}{p^2} \left(-g_{\mu\nu}+(1-\xi)\frac{p_{\mu}p_{\nu}}{p^2}\right) \nonumber \\
& \hspace{-20mm} \textrm{Tr}\left[(I \otimes v^{s_4}(p_4)u^{\dagger s_3}(p_3)) (\g^0\g^{\mu} \otimes \g^0\g^{\nu}) (u^{s_1}(p_1) v^{\dagger s_2}(p_2) \otimes I) \right]\,.
}
Finally, using the completeness relation for the photon polarization vectors
\eq{
\sum_{i} \e^{(i)}_{\mu}(p)\e^{*(i)}_{\nu}(p) = -g_{\mu\nu}+(1-\xi)\frac{p_{\mu}p_{\nu}}{p^2}
}
one arrives at
\ea{
\ma{M}_{e^+ e^- \to \mu^+ \mu^-} & = -\frac{ie^2}{p^2} \textrm{Tr}[(I \otimes v^{s_4}(p_4)u^{\dagger s_3}(p_3)) D_{\g}(p) \nonumber \\
& (u^{s_1}(p_1) v^{\dagger s_2}(p_2) \otimes I) ]
}
with
\eq{\label{propphoton}
D_{\g}(p) = \sum_{i} \Xi^{i}(p) \otimes \Xi^{i \dagger}(p)
}
and
\eq{\label{Xiphoton}
\Xi^{i}(p) = \e^{(i)}_{\mu}(p) \g^{\mu} \g^0, \quad \Xi^{\dagger i}(p) = \e^{*(i)}_{\mu}(p) \g^{\mu} \g^0 \,.
}
The expression $\slashed{\e}^{(i)}(p) \g^0$ is a particular case of the fermion propagator with $m=0$ and $k_{\mu} \to \e^{(i)}_{\mu}(k)$, so the eigenvalues and eigenvectors have the same form, subject to these specific values. In particular, one can directly use Eq.~(16) to obtain
\eq{\label{xiA}
\Xi^{i}(k) = M_{\e^{(i)}(k)} \sum_{i_1,i_2}\ket{i_1,i_2;\e^{(i)}(k)}\bra{i_1,i_2;\e^{(i)}(k)}
} 
which can be straightforwardly checked by direct inspection. Note that $\bra{i_1,i_2;\e^{(i)}(k)}$ does assume a conjugation of the vector components $\e^{(i)}_{\mu}(k)$. This quantity also appears naturally whenever real external photons are present in the interaction. For example, the lowest order Feynman diagram for Compton scattering gives the amplitude \cite{Schwartz:2013}
\eq{
\ma{M}_{e^- \g \to e^- \g} = e^2 \overline{u}(p_3) \slashed{\e}^*(p_4) \left(\frac{\slashed{k}+m}{s-m^2}\right) \slashed{\e}(p_2) u(p_1)
}
where $k=p_1+p_2$ and $s=\sqrt{k^2}$ is the center of mass energy. Using Eqs.~(\ref{rho}) and (\ref{Xiphoton}), this can be rearranged into the form
\eq{
\ma{M}_{e^- \g \to e^- \g} = \frac{4e^2 k_0}{s-m^2} u(p_3)^{\dagger} \Xi^{\dagger}(p_4) \check{\rho}(k) \Xi^{\dagger}(p_2) u(p_1)\,,
}
explicitely displaying the object $\Xi$.

Finally, we look at necessary but not sufficient conditions for the hermitian matrix $D_{\g}(p)$ to be positive definite and thus a physically valid density matrix. From Eq.~(\ref{propphoton}) one immediatly verifies that $D_{\g}(k)$ can be hermitian if all $\Xi^{i}(p)$ are hermitian. This is, however, a necessary but not sufficient condition, as the linear combination of tensor products in Eq.~(\ref{propphoton}) could still give an hermitian combination of non-hermitian matrices $\Xi^{i}(p) \otimes \Xi^{i \dagger}(p)$. Unfortunately, this general case is too complicated to be dealt with analytically. Focusing then on the matrices $\Xi^{i}(p)$, we can see form Eq.~(\ref{Xiphoton}) that all of the photons' polarizations must have only real components in order for them to be hermitian. The eigenvalues of $\Xi^{i}(p)$ are given by
\eq{
\e_0(p) \pm \sqrt{(\e_1(p))^2+(\e_2(p))^2+(\e_3(p))^2}
}
each with multiplicity 2. One can immediately see that, depending on the chosen gauge, two of the eigenvalues can be negative, which discards the hypothesis of $D_{\g}(k)$ being interpreted as a density matrix universally. Nevertheless, $\Xi^{i}(k)$ can have all eigenvalues positive if $\e_0(p)>0$ and $\e_0(p) > \sqrt{(\e_1(p))^2+(\e_2(p))^2+(\e_3(p))^2}$, which immediately implies that the tensor product $\Xi^{i}(k) \otimes \Xi^{i \dagger}(k)$ is also positive definite since its eigenvalues are the product of the individual matrices eigenvalues (which are all positive). Finally, since the sum of positive definite matrices is also a positive definite matrix, the matrix $D_{\g}(p)$ as defined by Eq.~(\ref{propphoton}) will also be positive definite in that case.

\section*{Section C: Proof of Eq.~(25)}

In this section we shall demonstrate a proof of Eq.~(25). Using the usual Feynman rules, one obtains the amplitude for the photon self-energy \cite{Schwartz:2013}
\begin{widetext}
\eq{
\ma{M}_{1-\textrm{loop}} = -e^2 \int \frac{d^4k}{(2\pi)^4} \e_{\mu}(p)\e^{*}_{\nu}(p) \textrm{Tr}\left[ \g^{\mu} \left(\frac{\slashed{p}-\slashed{k}+m}{(p-k)^2-m^2}\right) \g^{\nu} \left(\frac{\slashed{k}+m}{k^2-m^2}\right) \right]\,.
}
Using Eqs.~(16) and (23), we obtain
\ea{
\ma{M}_{1-\textrm{loop}} = &  -\frac{e^2}{16} \int \frac{d^4k}{(2\pi)^4} \frac{1}{(k^2-m^2)((p-k)^2)-m^2)} \textrm{Tr}\bigg[ \Xi(p) \sum_{\lambda_E,\lambda_s} (M_{p-k}+(-1)^{\lambda_{E'}}m) \ket{\lambda_{E'},\lambda_{s'};p-k}\bra{\lambda_{E'},\lambda_{s'};p-k} \nonumber \\
& \hspace{10mm} \Xi^{\dagger}(p) \sum_{\lambda_E,\lambda_s} (M_{k}+(-1)^{\lambda_{E}}m) \ket{\lambda_E,\lambda_s;k}\bra{\lambda_E,\lambda_s;k} \bigg] \nonumber \\
= & -e^2 \textrm{Tr}\left[(\Xi(p) \otimes \Xi^{\dagger}(p)) \check{\rho}_{2F}(p)\right]
}
with
\eq{
\check{\rho}_{2F}(p) = \int \frac{d^4k}{(2\pi)^4} \sum_{\substack{\la_{E},\la_{s}\\ \la_{E'},\la_{s'}}} (M_{p-k}+(-1)^{\lambda_{E'}}m) (M_{k}+(-1)^{\lambda_{E}}m) \ket{\la_{E'},\la_{s'}; p-k}\ket{\la_{E},\la_{s}; k} \bra{\la_{E},\la_{s}; k}\bra{\la_{E'},\la_{s'}; p-k}\,.
}
\end{widetext}
One may now use the relations $\ket{\la_{E},\la_{s}; k} = W_k \ket{\la_{E},\la_{s}}$, $\g^i \ket{\la_{E},\la_{s}} = -(-1)^{\la_E} (I \otimes \sigma_i)\ket{\la_{E},\la_{s}}$ (where $\si_i$ are the Pauli matrices) and $\g^0 \ket{\la_{E},\la_{s}} = (-1)^{\la_E}\ket{\la_{E},\la_{s}}$, to obtain the form
\eq{
\check{\rho}_{2F}(p) = R(p) S
}
with
\ea{
R(p) & = \int \frac{d^4k}{(2\pi)^4} \check{\rho}(p-k) \otimes \check{\rho}(k) \\
S & = \sum_{\substack{\la_{E},\la_{s}\\ \la_{E'},\la_{s'}}} \ket{\la_{E'},\la_{s'}}\ket{\la_{E},\la_{s}} \bra{\la_{E},\la_{s}}\bra{\la_{E'},\la_{s'}}
}
and where we have also used the identity $(A \otimes B)S = S(B \otimes A)$. This concludes the proof of Eq.~(20).

We now show that although $\check{\rho}_{2F}(p)$ is hermitian, it can never satisfy the positive semi-definiteness required of a physically valid density matrix. To show that it is hermitian, it is sufficient to see that
\ea{
\check{\rho}^{\dagger}_{2F}(p) & = \left(\left(\int \frac{d^4k}{(2\pi)^4} \check{\rho}(p-k) \otimes \check{\rho}(k)\right) S\right)^{\dagger} \nonumber \\
& = S^{\dagger}\left(\int \frac{d^4k}{(2\pi)^4} \check{\rho}^{\dagger}(p-k) \otimes \check{\rho}^{\dagger}(k)\right) \nonumber \\
& = S \left(\int \frac{d^4k}{(2\pi)^4} \check{\rho}(p-k) \otimes \check{\rho}(k)\right) \nonumber \\
& = \left(\int \frac{d^4k}{(2\pi)^4} \check{\rho}(p-k) \otimes \check{\rho}(k)\right) S \nonumber \\
& = \check{\rho}_{2F}(p) \nonumber \\
}
where the relations $S^{\dagger} = S$, $\check{\rho}^{\dagger}(k) = \check{\rho}(k)$, $(A \otimes B)S = S(B \otimes A)$ and a change of variables $k \to k-p$ was used. Finally, we show that $\check{\rho}_{2F}(p)$ will always have negative eigenvalues for all $p$, so it can never be positive semi-definite. Note that $R(p)^{\dagger}=R(p)$, as evidenced by the fact that $\check{\rho}_{2F}(p)$ is hermitian. Now define the matrix $C$ such that $C^2 = R(p)$. Taking the adjoint, we see that $C^{\dagger}C^{\dagger} = R(p)^{\dagger}$, which implies that $C^{\dagger} = \sqrt{R(p)^{\dagger}}$. On the other hand, $C=\sqrt{R(p)}$, so $C^{\dagger} = \sqrt{R(p)}^{\dagger}$, which means that $\sqrt{R(p)^{\dagger}} = \sqrt{R(p)}^{\dagger}$. Now define the eigenvalue spectrum of a matrix $A$ as $\la(A)$. For any two squared matrices $A$ and $B$, we have that $\la(AB) = \la(BA)$, since $AB$ and $BA$ have the same characteristic polynomial. In addition, we can use the later relation cyclically to obtain the result $\la(AB) = \la(\sqrt{A}B\sqrt{A})$. Using the later formula together with $\sqrt{R(p)^{\dagger}} = \sqrt{R(p)}^{\dagger}$, leads to the following relation regarding the eigenvalue spectrum of $\check{\rho}_{2F}(p)$:
\eq{\label{spectrumrho}
\la(\check{\rho}_{2F}(p)) = \la(\sqrt{R(p)}S\sqrt{R(p)}^{\dagger}) \,.
}
We can now use Sylvester's law of inertia, which states that if two matrices $A$ and $B$ are related by a third matrix $S$ through the form $A = SBS^{\dagger}$, then $A$ and $B$ share the same number of positive, negative and null eigenvalues. This implies that $\sqrt{R(p)}S\sqrt{R(p)}^{\dagger}$ and $S$ share the same number of negative eigenvalues. It is straightforward to show that the eigenvalues of $S$ are $\pm1$, each with a multiplicity of 8, therefore $\la(\sqrt{R(p)}S\sqrt{R(p)}^{\dagger})$ will share the same number of negative eigenvalues and, by Eq.~(\ref{spectrumrho}), so will $\la(\check{\rho}_{2F}(p))$. As a consequence, $\la(\check{\rho}_{2F}(p))$ will always have 8 negative eigenvalues and thus it can never be associated to a density matrix of a quantum state. This concludes the proof.\\

\section*{Section D: Renormalization effect on a virtual photon}

In this section we shall prove that renormalization does not physically alter the virtual photon operator of Eq.~(22). To begin, recall the renormalized form of the photon propagator \cite{Schwartz:2013}
\ea{
D^R_{\mu\nu}(k) & = \frac{-g_{\mu\nu}-k_{\mu}k_{\nu}/k^2}{k^2(1+C_1)} + \xi \frac{k_{\mu}k_{\nu}}{k^4} \\
& = \frac{1}{1+C_1}\frac{1}{k^2}\left(-g_{\mu\nu} + (\xi'-1) \frac{k_{\mu}k_{\nu}}{k^4} \right)
}
where $C_1$ is a finite number, corresponding to the renormalization correction, and $\xi' = (1+C_1)\xi$. We thus see that the only change in the propagator is a multiplication by a finite factor and a change of gauge, which have no consequence in the derivation of Eq.~(B5) in Section B. Consequently, the matrix elements of the operator $D_{\g}(p)$ are only multiplied by an overall factor.

\end{document}